\pdfoutput=1
\RequirePackage{ifpdf}

\documentclass[cits,11pt,a4paper]{article}
\usepackage{jinstpub}

\usepackage{graphicx}
\usepackage{natbib}

\usepackage{lineno}

\title{Electron Drift and Longitudinal Diffusion in High Pressure Xenon-Helium Gas Mixtures}

\collaboration{The NEXT Collaboration}
\author[1,a]{A.D.~McDonald,\note[a]{Corresponding author.}}
\author[1]{K.~Woodruff,}
\author[1]{B.~Al Atoum,}
\author[2]{D.~Gonz\'alez-D\'iaz,}
\author[1]{B.J.P.~Jones,}
\author[10]{C.~Adams,}
\author[17]{V.~\'Alvarez,}
\author[6]{L.~Arazi,}
\author[18]{I.J.~Arnquist,}
\author[4]{C.D.R~Azevedo,}
\author[19]{K.~Bailey,}
\author[20]{F.~Ballester,}
\author[17]{J.M.~Benlloch-Rodr\'{i}guez,}
\author[12]{F.I.G.M.~Borges,}
\author[17]{S.~C\'arcel,}
\author[17]{J.V.~Carri\'on,}
\author[21]{S.~Cebri\'an,}
\author[18]{E.~Church,}
\author[12]{C.A.N.~Conde,}
\author[2,14]{G.~D\'iaz,}
\author[17]{J.~D\'iaz,}
\author[5]{M.~Diesburg,}
\author[12]{J.~Escada,}
\author[20]{R.~Esteve,}
\author[17]{R.~Felkai,}
\author[11]{A.F.M.~Fernandes,}
\author[11]{L.M.P.~Fernandes,}
\author[14,8]{P.~Ferrario,}
\author[4]{A.L.~Ferreira,}
\author[11]{E.D.C.~Freitas,}
\author[14]{J.~Generowicz,}
\author[7]{A.~Goldschmidt,}
\author[14,8,b]{J.J.~G\'omez-Cadenas,\note[b]{NEXT Co-spokesperson.}}
\author[10]{R.~Guenette,}
\author[9]{R.M.~Guti\'errez,}
\author[10]{J.~Haefner,}
\author[19]{K.~Hafidi,}
\author[3]{J.~Hauptman,}
\author[11]{C.A.O.~Henriques,}
\author[2]{J.A.~Hernando~Morata,}
\author[14,17]{P.~Herrero,}
\author[20]{V.~Herrero,}
\author[19]{S.~Johnston,}
\author[17]{M.~Kekic,}
\author[16]{L.~Labarga,}
\author[1]{A.~Laing,}
\author[5]{P.~Lebrun,}
\author[17]{N.~L\'opez-March,}
\author[9]{M.~Losada,}
\author[11]{R.D.P.~Mano,}
\author[10]{J.~Mart\'in-Albo,}
\author[14]{A.~Mart\'inez,}
\author[2,17]{G.~Mart\'inez-Lema,}
\author[14]{F.~Monrabal,}
\author[11]{C.M.B.~Monteiro,}
\author[20]{F.J.~Mora,}
\author[17]{J.~Mu\~noz Vidal,}
\author[17]{P.~Novella,}
\author[1,c]{D.R.~Nygren,\note[c]{NEXT Co-spokesperson.}}
\author[17]{B.~Palmeiro,}
\author[5]{A.~Para,}
\author[17,d]{J.~P\'erez,\note[d]{Now at Laboratorio Subterr\'aneo de Canfranc, Spain.}}
\author[17]{M.~Querol,}
\author[17]{J.~Renner,}
\author[19]{J.~Repond,}
\author[19]{S.~Riordan,}
\author[15]{L.~Ripoll,}
\author[9]{Y.~Rodriguez Garcia,}
\author[20]{J.~Rodr\'iguez,}
\author[1]{L.~Rogers,}
\author[14]{B.~Romeo,}
\author[17]{C.~Romo-Luque,}
\author[12]{F.P.~Santos,}
\author[11]{J.M.F. dos~Santos,}
\author[6]{A.~Sim\'on,}
\author[13,e]{C.~Sofka,\note[e]{Now at University of Texas at Austin, USA.}}
\author[17]{M.~Sorel,}
\author[13]{T.~Stiegler,}
\author[20]{J.F.~Toledo,}
\author[14]{J.~Torrent,}
\author[17]{A.~Us\'on,}
\author[4]{J.F.C.A.~Veloso,}
\author[13]{R.~Webb,}
\author[6]{R.~Weiss-Babai,}
\author[13,f]{J.T.~White,\note[f]{Deceased.}}
\author[17]{N.~Yahlali}
\emailAdd{austin.mcdonald@uta.edu}
\affiliation[1]{
Department of Physics, University of Texas at Arlington, Arlington, TX 76019, USA}
\affiliation[2]{
Instituto Gallego de F\'isica de Altas Energ\'ias, Univ.\ de Santiago de Compostela, Campus sur, R\'ua Xos\'e Mar\'ia Su\'arez N\'u\~nez, s/n, Santiago de Compostela, E-15782, Spain}
\affiliation[3]{
Department of Physics and Astronomy, Iowa State University, 12 Physics Hall, Ames, IA 50011-3160, USA}
\affiliation[4]{
Institute of Nanostructures, Nanomodelling and Nanofabrication (i3N), Universidade de Aveiro, Campus de Santiago, Aveiro, 3810-193, Portugal}
\affiliation[5]{
Fermi National Accelerator Laboratory, Batavia, IL 60510, USA}
\affiliation[6]{
Nuclear Engineering Unit, Faculty of Engineering Sciences, Ben-Gurion University of the Negev, P.O.B. 653, Beer-Sheva, 8410501, Israel}
\affiliation[7]{
Lawrence Berkeley National Laboratory (LBNL), 1 Cyclotron Road, Berkeley, CA 94720, USA}
\affiliation[8]{
Ikerbasque, Basque Foundation for Science, Bilbao, E-48013, Spain}
\affiliation[9]{
Centro de Investigaci\'on en Ciencias B\'asicas y Aplicadas, Universidad Antonio Nari\~no, Sede Circunvalar, Carretera 3 Este No.\ 47 A-15, Bogot\'a, Colombia}
\affiliation[10]{
Department of Physics, Harvard University, Cambridge, MA 02138, USA}
\affiliation[11]{
LIBPhys, Physics Department, University of Coimbra, Rua Larga, Coimbra, 3004-516, Portugal}
\affiliation[12]{
LIP, Department of Physics, University of Coimbra, Coimbra, 3004-516, Portugal}
\affiliation[13]{
Department of Physics and Astronomy, Texas A\&M University, College Station, TX 77843-4242, USA}
\affiliation[14]{
Donostia International Physics Center (DIPC), Paseo Manuel Lardizabal, 4, Donostia-San Sebastian, E-20018, Spain}
\affiliation[15]{
Escola Polit\`ecnica Superior, Universitat de Girona, Av.~Montilivi, s/n, Girona, E-17071, Spain}
\affiliation[16]{
Departamento de F\'isica Te\'orica, Universidad Aut\'onoma de Madrid, Campus de Cantoblanco, Madrid, E-28049, Spain}
\affiliation[17]{
Instituto de F\'isica Corpuscular (IFIC), CSIC \& Universitat de Val\`encia, Calle Catedr\'atico Jos\'e Beltr\'an, 2, Paterna, E-46980, Spain}
\affiliation[18]{
Pacific Northwest National Laboratory (PNNL), Richland, WA 99352, USA}
\affiliation[19]{
Argonne National Laboratory, Argonne, IL 60439, USA}
\affiliation[20]{
Instituto de Instrumentaci\'on para Imagen Molecular (I3M), Centro Mixto CSIC - Universitat Polit\`ecnica de Val\`encia, Camino de Vera s/n, Valencia, E-46022, Spain}
\affiliation[21]{
Laboratorio de F\'isica Nuclear y Astropart\'iculas, Universidad de Zaragoza, Calle Pedro Cerbuna, 12, Zaragoza, E-50009, Spain}

\usepackage{microtype}
\usepackage{rotating}
\usepackage{booktabs}
\usepackage{multirow}
\usepackage{tabularx}
\usepackage{dcolumn}
\usepackage{bm}
\usepackage[version=3]{mhchem}

\newcolumntype{R}{>{\raggedleft\arraybackslash}X}
\newcolumntype{Y}{>{\centering\arraybackslash}X}

\makeatletter
\g@addto@macro\bfseries{\boldmath}
\makeatother

\abstract{We report new measurements of the drift velocity and longitudinal diffusion coefficients of electrons in pure xenon gas and in xenon-helium gas mixtures at 1-9 bar and electric field strengths of 50-300 V/cm.  In pure xenon we find excellent agreement with world data at all $E/P$, for both drift velocity and diffusion coefficients.   However, a larger value of the longitudinal diffusion coefficient than theoretical predictions is found at low $E/P$ in pure xenon, below the range of reduced fields usually probed by TPC experiments. A similar effect is observed in xenon-helium gas mixtures at somewhat larger $E/P$. Drift velocities in xenon-helium mixtures are found to be theoretically well predicted.  Although longitudinal diffusion in xenon-helium mixtures is found to be larger than anticipated, extrapolation based on the measured longitudinal diffusion coefficients suggest that the use of helium additives to reduce transverse diffusion in xenon gas remains a promising prospect.}

\keywords{Gaseous detectors, Noble element TPCs}

\begin{document}
\maketitle

\section{Introduction}

High pressure gas time projection chambers (TPCs) are an important tool for studying low-energy particle interactions
because of their excellent energy resolution and strong topological reconstruction capabilities. The precise reconstruction
of interaction products and radioactive decays in gas TPCs has been utilized for
efficient signal selection and background suppression in many applications~\cite{Gonzalez-Diaz:2017gxo}. 

The NEXT HPXe TPC experiment~\cite{alvarez2013initial,alvarez2013operation,alvarez2013near,Ferrario:2015kta}, for example, uses high pressure (10-15~bar) xenon gas TPCs to search for neutrinoless double beta decay of $^{136}$Xe.  The signature of this ultra-rare process is an event reconstructed with energy consistent with the Q-value of the decay, a topology of two back-to-back electrons, identified via the high dE/dx these produce at the track ends. Extrapolated resolutions of better than 1\% FWHM at the Q-value have been demonstrated in the NEXT-NEW detector~\cite{Renner:2018ttw}, and the power of topological reconstruction has been studied and found to be consistent with expectations from simulations~\cite{Ferrario:2015kta}. Other alternative molecular species such as CO$_{2}$-, CH$_{4}$- and CF$_4$-doped in xenon at atmospheric pressure have been recently studied, with electroluminescence yield and its associated statistical fluctuations presented in~\cite{henriques2019electroluminescence, henriques2017secondary}.

The spatial resolution of a TPC depends on two factors: 1) granularity of the tracking plane, and 2) charge spreading through diffusion as electrons drift in the gas.  For longer drift distances, the challenges of diffusion become more pronounced, with the size of the arriving charge distribution scaling like the square root of the drift distance, given any specific gas or mixture.  As detectors become larger in size, the effects of diffusion thus become a more significant concern.  Electron drift properties in several
noble gases at low pressure have been studied
in~\cite{pack1962drift,pack1992longitudinal,koizumi1986momentum,bowe1960drift}, and at high pressure in xenon gas
~\cite{hunter1988low,alvarez2013near,alvarez2013ionization,lorca2014characterisation,Simon:2018vep}. Other notable work has been undertaken in noble gases at low pressure in~\cite{patrick1991electron,english1953grid}, in hydrogen-doped xenon at high pressure in~\cite{kobayashi2006ratio}, and in Xenon-TMA mixtures at high pressure~\cite{Gonzalez-Diaz:2015oba,Nakajima:2015meb}.

\begin{figure}
\begin{centering}
\includegraphics[width=0.49\columnwidth]{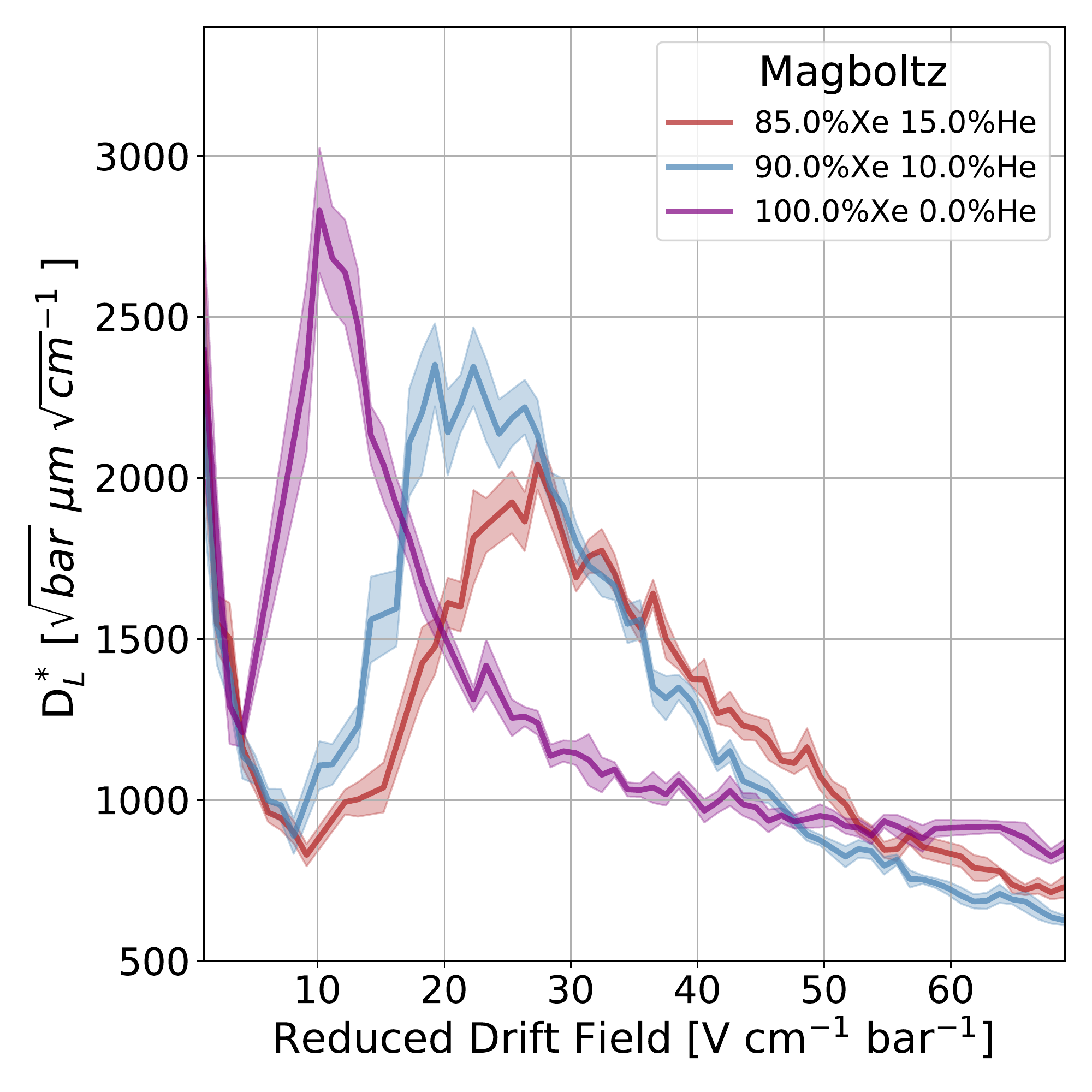}
\includegraphics[width=0.49\columnwidth]{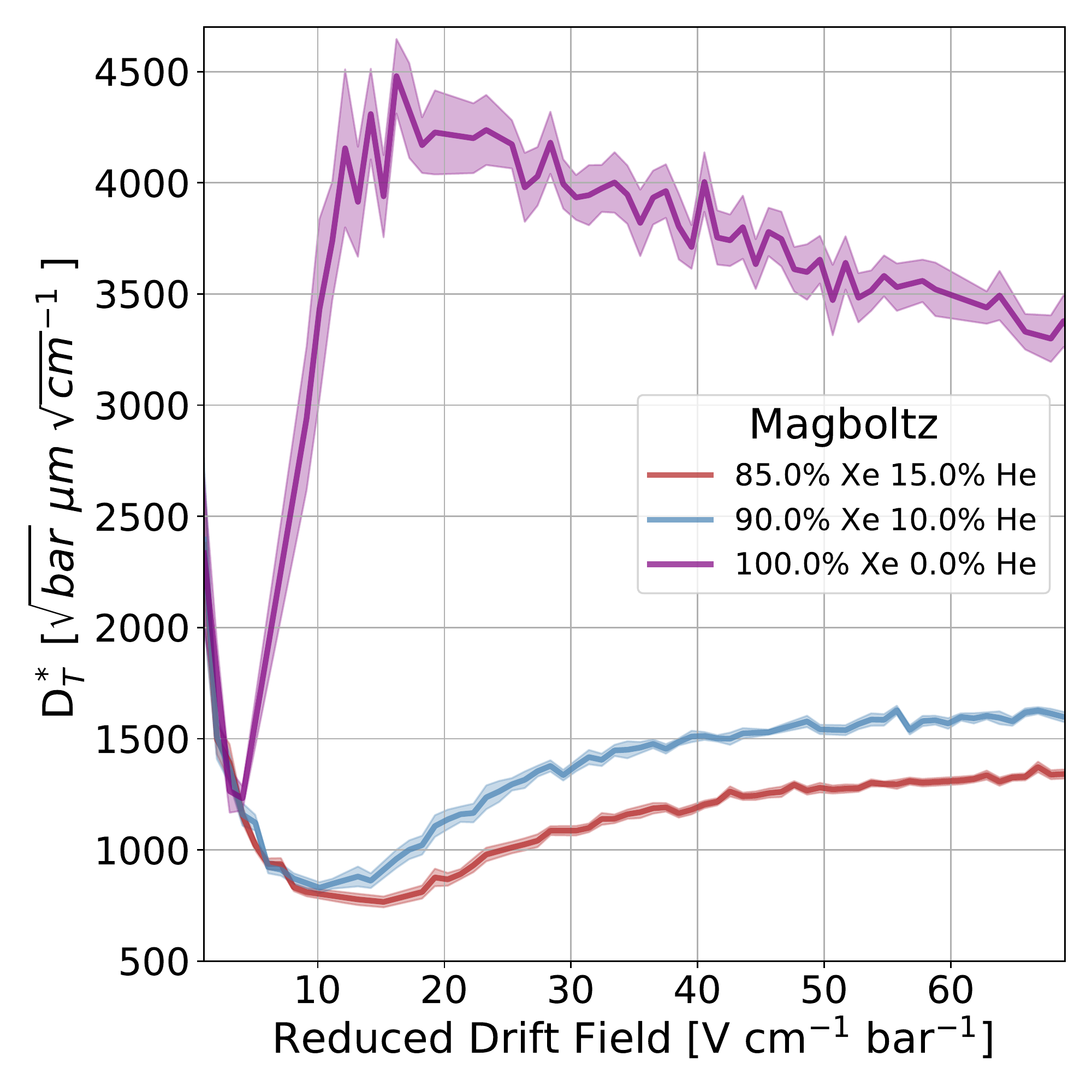}
\par\end{centering}
\caption{MagBoltz predictions of the dependence of the reduced longitudinal (left) and transverse (right) diffusion coefficients(equation \ref{Dstar}) on helium concentration. The line represents the exact value from the simulation and the shaded region is the error in the simulation. 
 \label{fig:MagboltzPredictions}}
\end{figure}

Electron drift properties in mixtures of noble elements at high pressures are less well studied.  In this work we present novel measurements in gas mixtures that are of interest to the NEXT collaboration: xenon with a small admixture of helium.   It has been noted that the addition of small concentrations of helium to high-pressure xenon is predicted to reduce transverse
electron diffusion, while imposing only a relatively small effect on longitudinal diffusion~\cite{XePa,Felkai:2017oeq}.  Plots demonstrating this predicted effect, which derives from the cooling of electrons by elastic scattering off relatively light (compared to xenon) helium atoms, are shown in Fig~\ref{fig:MagboltzPredictions}. These predictions are obtained from the {\tt MagBoltz} software package, which we use extensively throughout this work.

The relevant concentrations of helium to achieve this effect are in the 5-20\% by volume.  Although these cooling effects appear to present a promising way to improve the spatial resolution of high pressure gas detectors, they have yet to be experimentally verified.  

This paper presents measurements of electron drift velocity and longitudinal diffusion coefficients in xenon gas and xenon-helium gas mixtures over a range of pressures and electric field strengths covering the space of $E/P$ (reduced drift field) from 5.5 V/cm-bar to 300 V/cm-bar.  

The layout of the paper is as follows.   Sec.~\ref{sec:Methodology} describes our methodology, including the gas mixing protocols (Sec.~\ref{sec:GasMixing}), apparatus used in this work (Sec.~\ref{sec:Aparatus}),  data analysis methods (Sec.~\ref{sec:Analysis}) and studies of systematic uncertainties (Sec.~\ref{sec:Systematics}). We then present results in Sec.~\ref{sec:Results}.  We first present measured velocities and diffusion coefficients in pure xenon gas in Sec.~\ref{sec:pure}. This is followed by electron drift velocity and longitudinal diffusion in xenon doped with 10\% and 15\% helium in Sec.~\ref{sec:xehe}. Finally we present a discussion of these results and their implications in Sec.~\ref{sec:Discussion}.

\section{Methodology}
\label{sec:Methodology}
\subsection{Gas Handling}

\label{sec:GasMixing}

\begin{figure}[b!]
\begin{centering}
\includegraphics[width=0.34\columnwidth]{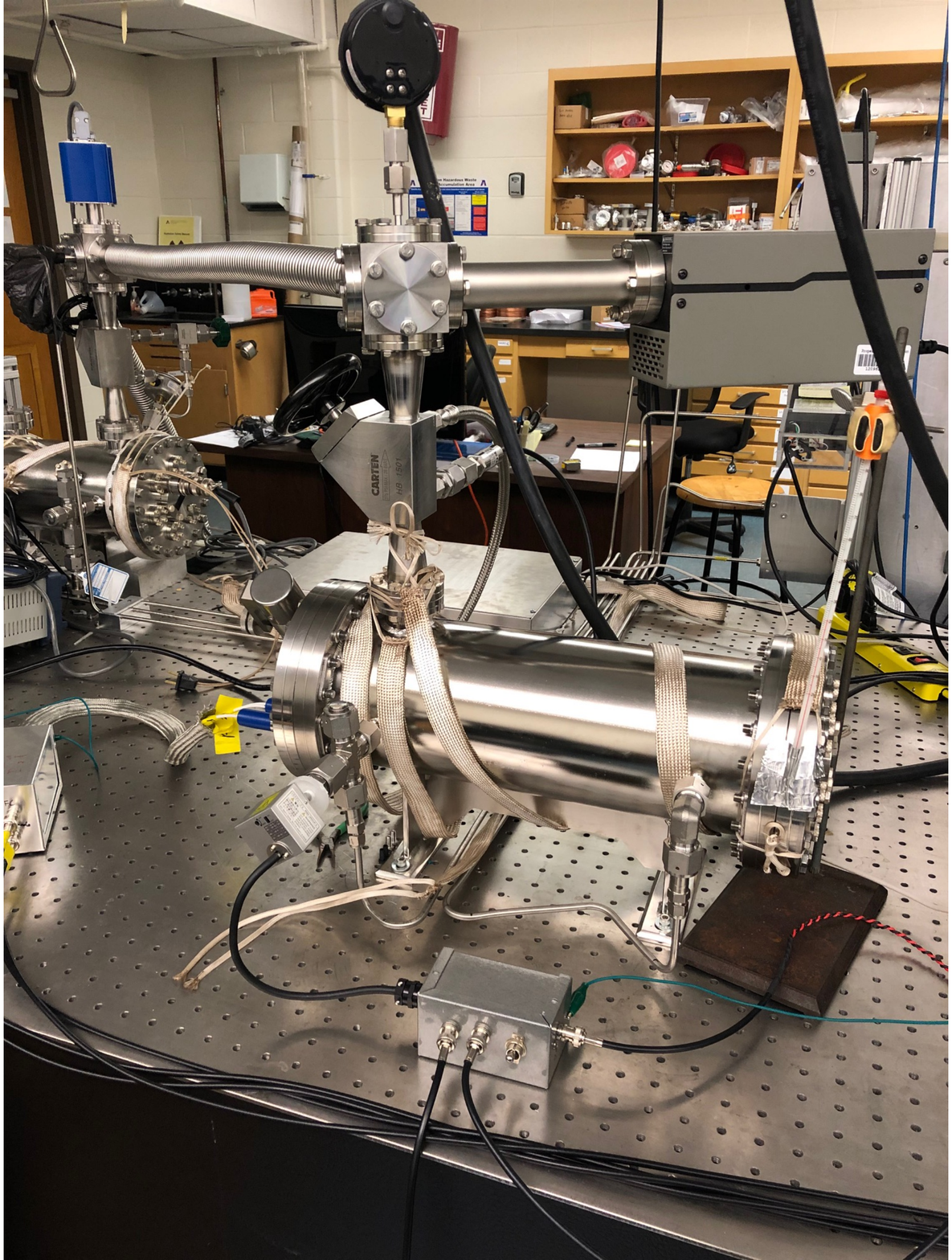}
\includegraphics[width=0.65\columnwidth]{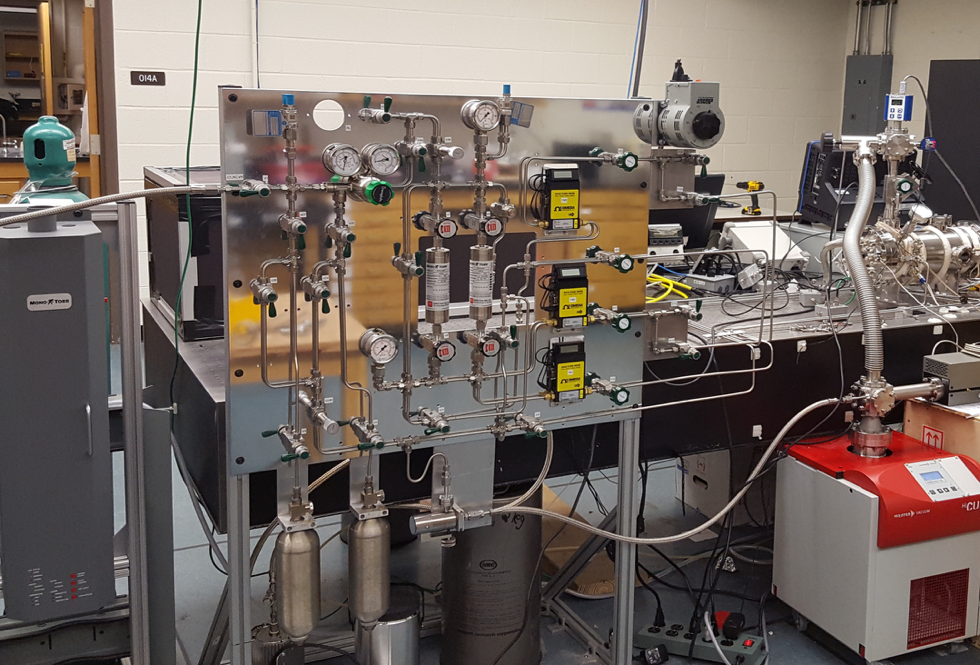}
\par\end{centering}
\caption{Left: Photograph of the 5 litre pressure vessel with the electron drift apparatus inside. Right: Photograph of the gas system that supports this vessel and another similar one, used as a holding volume for gas in this study.\label{fig:GasSystemPhoto}}
\end{figure}

The work in this paper was undertaken using two 5~litre pressure vessels  sharing a common circulation system in the High Pressure Xenon Gas Laboratory at the University of Texas at Arlington.  The vessel housing the electron drift apparatus is shown in Fig.~\ref{fig:GasSystemPhoto}, left, and a photograph of the gas-handling system in Fig.~\ref{fig:GasSystemPhoto}, right.

The system was constructed with all Swagelok parts and 1/4 inch stainless steel tubing. A MetroCad PumpWorks {\tt PW-2070} pump circulates gas through the system in a closed loop at pressures between 1 and 10~bar.  Flow is measured on the inlet and outlet of the vessel via mass flow meters, with typical flow rates between one and ten litres per minute.  The gases used in this study are all Ultra-High Purity grade from AirGas, which carry few ppm level specifications for oxygen, nitrogen, water and carbon dioxide impurity.  The system is filled at low flow rate through a MonoTorr {\tt PS-4-MT3} hot getter, which is also in the circulation path and cleans the gas to a specification of $\leq$ 1~ppb in oxygen, water and nitrogen in xenon, argon and helium carrier gases.  Pressure is monitored coarsely with analogue gauges and more accurately, to 10~mbar precision, with a pressure transducer which is calibrated using set points at vacuum and 9 atm. Burst disks protect the system from over-pressurization.

Gases with boiling points below that of nitrogen (in particular xenon) can be recaptured after study by submerging a collection cylinder in liquid nitrogen. The gases in the system condense into the cylinder, which can then be closed and is allowed to return to room temperature through ambient heating.

The full system was rigorously tested for leaks and has a leak rate of $<1\times 10^{-9}$ cc He $s^{-1}$, the lower limit of our helium leak-checker.  The gas system is coupled to the pressure vessels in a closed loop of permanent tube connections.  The full system including both vessels and gas panel can be evacuated to better than  $1\times10^{-5}$~mbar, which is a standard procedure before adding any new gas into the device.  After opening to atmosphere, a much longer pump and bake out is performed to drive the vacuum to better than $1\times10^{-6}$~mbar.  The room temperature was monitored throughout the measurements period and showed no significant deviation from its initial value of 21 $^{\circ}$C during the study.

The mixing and handling process of gases in this study was devised to provide accurate and consistent concentrations while maintaining economy given the expense of xenon gas. Pure xenon was added to the vessel, and measurements were made during the first fill at 1, 3, 6, and 9 bar. Once 9 bar was reached, helium was added to 10\% concentration, and measurements were performed. After completion, a vent line into the second, evacuated vessel was opened and the pressure was reduced by flowing gas into the lower pressure volume. Then the original volume was resealed, the xenon in the vented gas mixture was recaptured to the extent possible, and the residual (around 100~mbar) was evacuated. This procedure was repeated three times, probing four pressures. The pressures were accurately recorded and used in the derivation of $E/P$ and all experimentally derived quantities.  On some plots the reported labels are rounded to the nearest bar for brevity, the data will be included as a supplementary file~\footnote{\url{http://www-hep.uta.edu/~bjones/XenonHelium/}}. 

Once the last measurement was made on the mixture, the recaptured solid xenon was pumped to remove any residual helium gas.  Monitoring on a residual gas analyzer at the vacuum pump demonstrated that helium was removed effectively. After removing the helium trace, the procedure was repeated at the next concentration point, of 15\% helium in xenon.

\begin{figure}
\begin{centering}
\includegraphics[width=0.9\columnwidth]{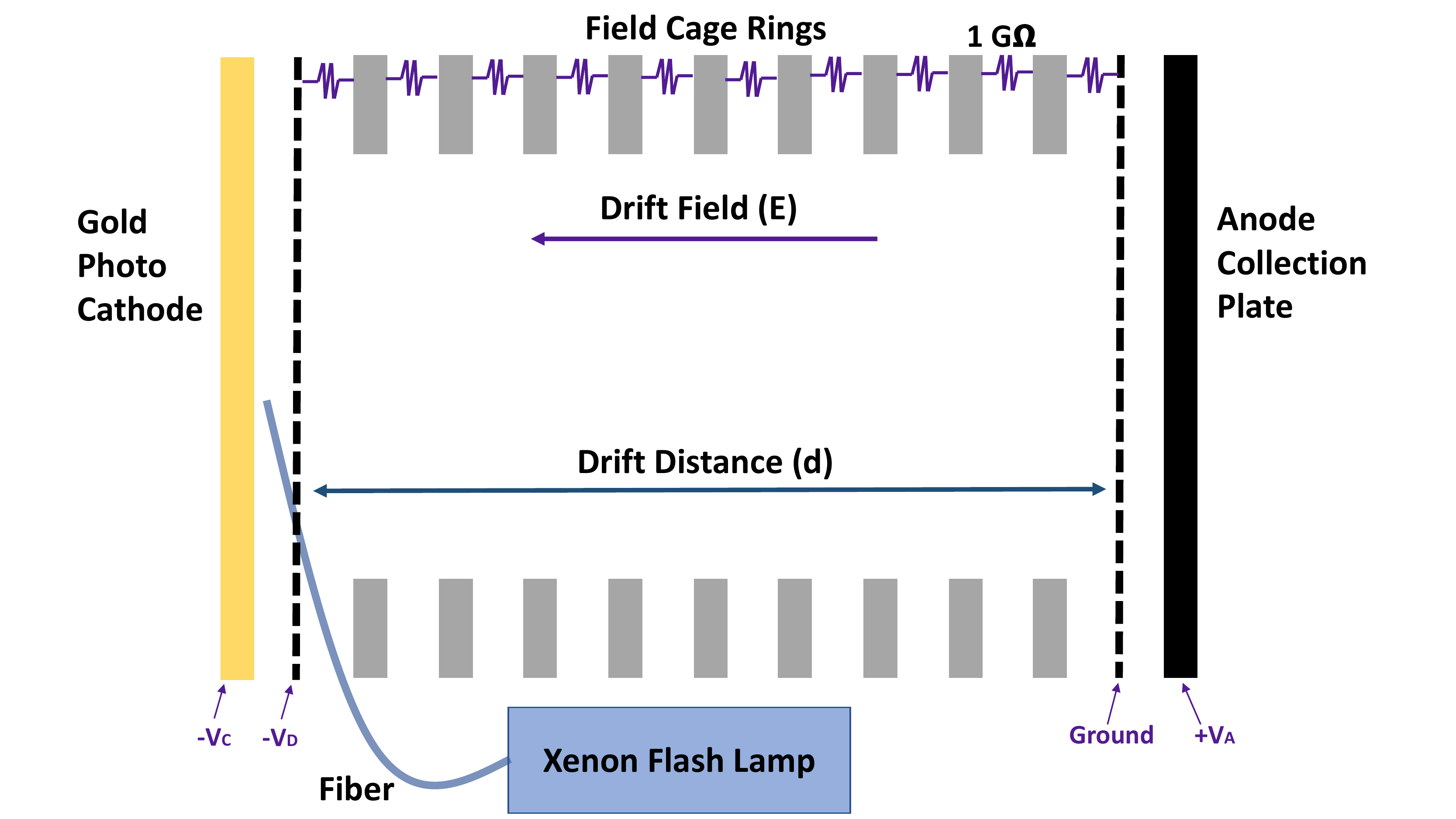}
\par\end{centering}
\caption{Schematic of the electron drift cell.\label{fig:schematic}}
\end{figure}

\subsection{Electron Production, Drift and Collection}
\label{sec:Aparatus}
The measurement of electron transport properties uses a gridded drift cell, shown schematically in Fig~\ref{fig:schematic}. Electrons are liberated from a gold photo-cathode by a pulsed xenon flash lamp, delivered by an optical fiber directed at the gold surface.
The setup is somewhat similar to that used by others to study electron transport in liquid argon~\cite{Li:2015rqa}, with some important differences.
Unlike in the case of liquid argon experiments where electrons are freed into a conduction band of the liquid, in gas, electrons thermalizing near the photo-cathode surface are more likely to back-scatter and be reabsorbed into the conductor by their image charges~\cite{coelho2007measurement,borghesani2011injection,smejtek1973hot}. In order to overcome this phenomenon an extraction field is applied on the photo cathode to pull the electrons away from it.  Our apparatus thus includes an additional drift stage with elevated electric field to extract electrons from the gold. The extraction region was made between the photo-cathode and a custom-made hexagonal mesh with 2.5~mm diameter holes and 5~mil(127$\mu$m) lands, photo-etched in 3 mil(76.2$\mu$m) thickness stainless steel. The extraction region is 0.39~cm long and biased to a constant electric field of 300~V/cm for all tests.  For drift fields lower than this value, a significant fraction of electrons are lost during transmission from the higher-field extraction region to the lower-field drift region.  A scan of the extraction field strength demonstrated the expected variation in transmitted signal amplitude due to electron transparency of the mesh, but introduced no observable variation into the measured electron transport parameters. Likewise, variations in the photoelectron extraction efficiency also do not introduce variations in the measured electron transport parameters.

\begin{figure}
\begin{centering}
\includegraphics[width=0.99\columnwidth]{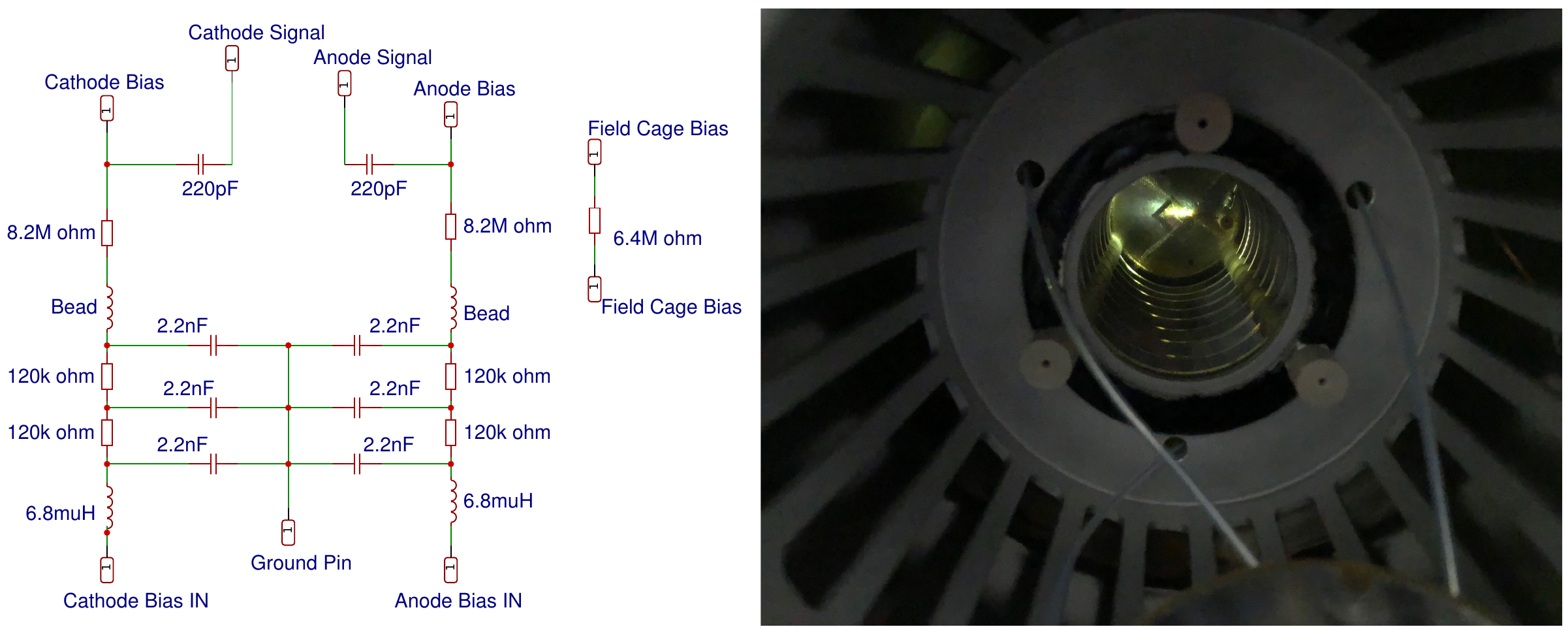}
\par\end{centering}
\caption{Left: Schematic of the electron drift circuit; Right; view into drift region from the collection plane upward. Light is being shone through the UV transmitting fiber and illuminating the gold photo-cathode.\label{fig:circuitandgold}}
\end{figure}

A xenon flash lamp (Hamamatsu model {\tt L13651-12}), equipped with a quartz window for the transmission of the UV light, was used as a light source. The lamp was coupled to an optical feedthrough that was made in-house from quartz fiber and potted in Stycast 2080 vacuum epoxy. The feedthrough was pressure and leak tested before use, and checked for optical continuity before closing the vessel.  A photograph showing the fiber shining onto the gold photo-cathode is shown in Fig.~\ref{fig:circuitandgold}, right. 
 
The drift region is 14.12~cm long and the field uniformity is maintained by 3-inch stainless steel field shaping rings separated by 1.25~cm ceramic spacers. The voltage from the top of the field cage is stepped down by 1~G$\Omega$ SlimMox Ohmite resistors, and terminated at ground at the bottom of the cage where a second mesh is mounted. The collection plate, where electrons are detected, is located 0.16~cm behind the grounded mesh, and is comprised of a copper-clad FR4 circuit board connected to a pin feed-through by copper wire.  The collection field between this plate and the grounded mesh was held at 3125~V/cm to accelerate electrons quickly across the final gap and avoid inductive effects that would bias the time measurements.  Electronic pulses are recorded from both the gold photo cathode and copper collection plate using a capacitive coupling.  The circuit diagram for the readout system is shown in Fig.~\ref{fig:circuitandgold}, left. The anode and cathode biases are filtered with IC and RC filters to minimize switching noise from the power supplies. The signals are then measured with an RC circuit with a time constant of $1.8 ms$. The field cage bias is not filtered and is terminated at ground inside the vessel with a mechanical connection to the vessel body.

The flash lamp was pulsed every 53~ms, chosen to ensure the electron pulses were out of phase with any line and power supply switching noise. This was used as the trigger since the lamp flash time is $< 500ns$ no uncertinty was added.  For each run, 1000 waveforms from both the photo cathode and collection plate were recorded on a LeCroy {\tt HDO6104} for each electric field and pressure.  Measurements were made at various electric fields in the drift region ranging from 50~V/cm to 300~V/cm.

\subsection{Analysis Methodology}
\label{sec:Analysis}
An electron swarm in a static electric field can be treated according to the laws of classical electromagnetism and hydrodynamics. The charge carrier density $n(\vec{r},t)$ in a strong, uniform electric
field is governed by the diffusion equation, which states that, in
the absence of spacecharge effects, attachment, gain or chemical reactions \cite{ViehlandBook}:

\begin{equation}
\frac{\partial}{\partial t}n(\vec{r},t)+v_{d}\frac{\partial}{\partial z}n(\vec{r},t)-D_{T}\left(\frac{\partial^{2}}{\partial x^{2}}+\frac{\partial^{2}}{\partial y^{2}}\right)n(\vec{r},t)-D_{L}\frac{\partial^{2}}{\partial z^{2}}n(\vec{r},t)=0.
\end{equation}Here the electric field is taken to lie in the $\hat{z}$ direction,
$v_{d}$ is the drift velocity and $D_{L}$ and $D_{T}$ are the longitudinal and transverse diffusion coefficients, respectively, which all depend on the gas mixture, pressure,
and electric field. An analytic solution to this equation can be
found, given an initial thin disk of electrons with Gaussian profile in
the z direction of width $\sigma_{0}$, and radius $r_{0}$. Integrating
over the transverse coordinates yields the longitudinal charge density
in the $\hat{z}$ direction as a function of time:

\begin{equation}
j(z_{1},t)=X\frac{(z_{1}+v_{d}t)D_{L}+\sigma_{0}^{2}v_{d}^{2}}{\left(\sigma_{0}^{2}+2D_{L}t\right)^{3/2}}\left[1-\exp\left(-\frac{r_{0}^{2}}{4D_{T}t}\right)\right]\exp\left[-\frac{(z_{1}-v_{d}t)^{2}}{2(\sigma_{0}^{2}+2D_{L}t)}\right].\label{eq:FullProfile}
\end{equation}

We have collected some constants into an uninteresting numerical normalization pre-factor, $X$. If we set $z_{1}$ equal to the drift distance $d$ then
this expression gives the flux of electrons arriving at our detection plane. In that case, the
term on the right only gives significant contributions in the time
window around the ``drift time'', $t_{d}=d/v_{d}$, with effective width
$\sigma_t$:

\begin{equation}
\sigma_t=\sqrt{\sigma_{0(t)}^{2}+2D_{L(t)}t_{d}}.\label{eq:SigmaDef}
\end{equation}
In the above we defined two new constants, $\sigma_{0(t)}=\sigma_{0}/v_{d}$ and $D_{L(t)}=D_{L}/v_{d}^{2}$ that effectively
map $\sigma_{0}$ and $D_{L}$ into the time domain.

If the width $\sigma_t$ is sufficiently narrow then the left two factors
vary little over this window and they can be assumed to be constant over
the pulse duration. This approximation yields a Gaussian profile centered
around the drift time:

\begin{figure}
\begin{centering}
\includegraphics[width=0.99\columnwidth]{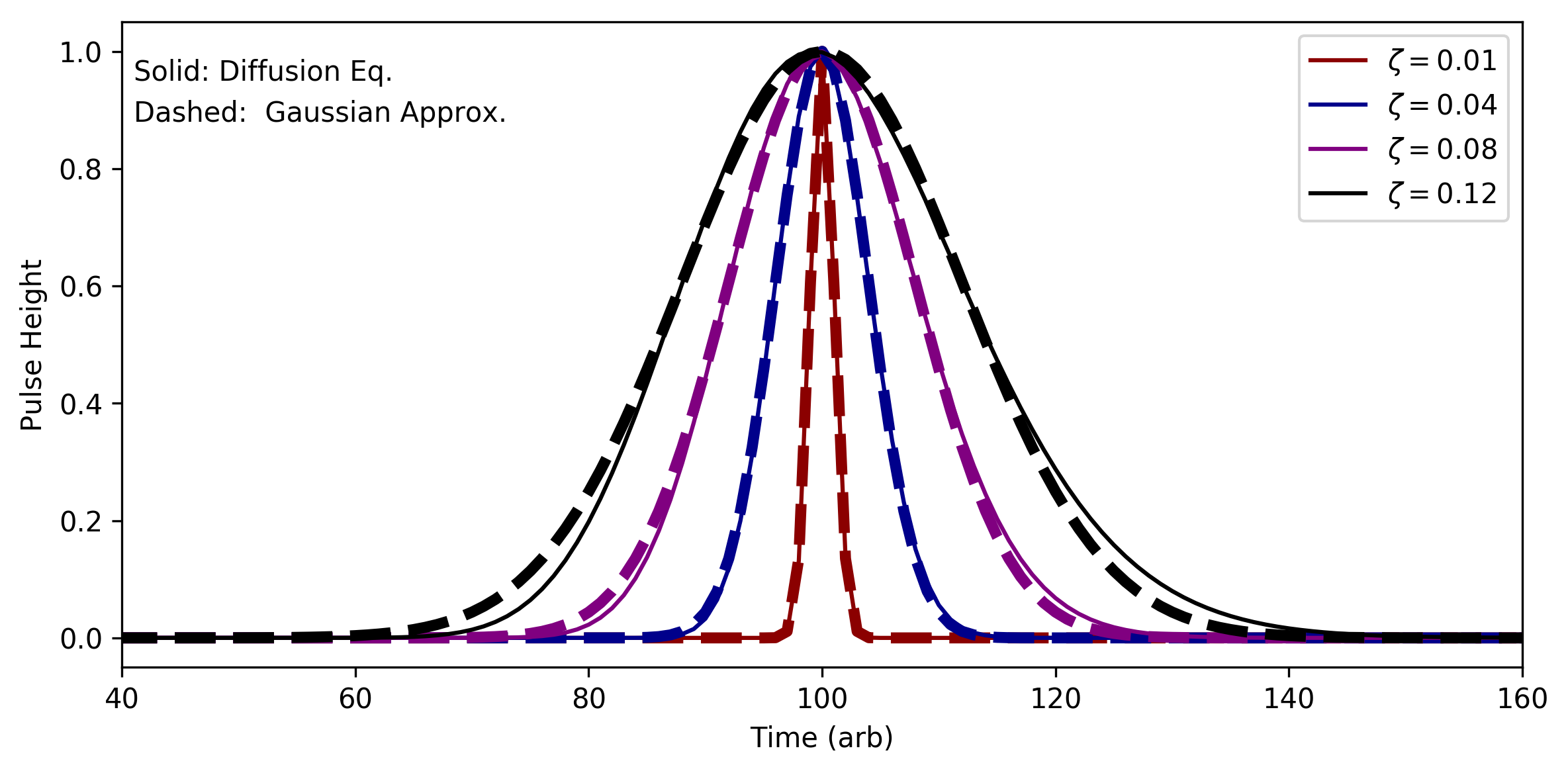}
\par\end{centering}
\caption{Comparison of the Gaussian approximation to the full diffusion equation solution. In this study, the relevant range of values of $\zeta=\sigma_t/t_{d}=\sqrt{2D_{t}/t_{d}}$ are between 0.01 and 0.12. \label{fig:DiffApprox}}
\end{figure}

\begin{equation}
j(t)\rightarrow\frac{N}{\sqrt{2\pi\sigma^{2}}}\exp\left[-\frac{(t-t_{d})^{2}}{2\sigma^{2}}\right],\label{eq:GaussianProfile}
\end{equation}where N is the total number of electrons arriving. This approximation
will be accurate as long as the drift time,  $t_d$,  is suitably long relative to the
pulse width in time, $\sigma_{0(t)}$. Comparison of Eq.~\ref{eq:FullProfile} with
\ref{eq:GaussianProfile} is shown in Fig.~\ref{fig:DiffApprox} for various ratios $\zeta=\sigma_t/t_{d}=\sqrt{2D_{t}/t_{d}}$,
given an initially negligible $\sigma_{0}$, which provides the worst-case
scenario. The range of pulse widths in these studies is $0.005\leq\zeta\leq0.04$,
with the majority being in the range $0.01\leq\zeta\leq0.02$. Use
of the Gaussian approximation is found to be a sub-dominant source
of error in our measurements compared to other systematic uncertainties.

In practice, the drift velocity is determined by the ratio of the known drift distance $d$ and the measured drift time $t_d$. To obtain $D_L$, the relationship \ref{eq:SigmaDef} is inverted to map a measured value for $\sigma_t$ onto a measurement of the longitudinal diffusion coefficient.  

\begin{equation}
D_{L(t)}=\frac{\sigma_t^{2}-\sigma_{0(t)}^{2}}{2t_{d}}.
\end{equation}

The initial width $\sigma_{0(t)}$ above represents the width of the flash lamp pulse, and any irreducible width inherent in the system readout electronics. It is determined by operating the system in vacuum conditions, where the electrons move ballistically and effectively arrive instantaneously at the anode. A measured value of  $\sigma_{0(t)}=0.1235\pm0.0004 \mu s$ is obtained, with the very small uncertainty deriving from repeating the measurement in different field configurations and taking the standard deviation.  
 
Although we do not measure $D_T$, theoretically the transverse diffusion can then be determined from the Generalized Einstein Relations: ~\cite{robson1972thermodynamic}
\begin{equation}
    \frac{D_L}{D_T} = 1 + (1+\Delta)\frac{E}{\mu}\frac{\partial{\mu}}{\partial{E}} \,,\label{eq:GenEinsteinRel}
\end{equation}
where $E$ is the electric field strength, and $\mu$ is the electron mobility defined by $\mu = \frac{v_d}{E}$. The Wannier relation is obtained in the limit of massless electrons and negligible inelasticity and corresponds to the special case $\Delta=0$:
\begin{equation}
    \frac{D_L}{D_T} = \frac{1}{v_d}\frac{E}{P}\frac{\partial{v_d}}{\partial{(E/P)}} \,,\label{eq:WannierRel}
\end{equation}
Inspection of simulations suggests that for the majority of gases and mixtures in the present work, this approximation is only applicable over a limited range of $E/P$. This will be discussed further in Sec.~\ref{sec:Discussion}

The detected waveform shape is influenced by the initial electron pulse width, which primarily drives the rising edge, and the decay time of the RC which primarily drives the falling edge, and the inductive current for electrons crossing the gap, which effectively broadens the pulse.   The expected pulse shape in the general case is a Gaussian function of charge arrival with width $\sigma_t$, convolved with a top-hat function for the inductive current of width $w$, convolved with an exponential decay of the electronics response with lifetime $\tau$.  To a good approximation in our regime of interest the first convolution can be well approximated by a Gaussian of width ${\sigma'}_t^2=\sigma_t^2+(2/3 w)^2$ and center ${t'}_0=t_0+w/2$.  The second convolution can performed analytically and yields the functional form~\cite{Li:2015rqa}: 
\begin{equation}
    f(t,{t'}_0,{\sigma'}_t, \tau, a, c)  = a\cdot e^{\frac{{\sigma'}_t^2-2(t-t'_0)\tau}{2\tau^2}} \cdot \bigg[ 1+ \textrm{Erf} \bigg( \frac{-{\sigma'}_t^2+(t-{t'}_0)\tau}{\sqrt{2}\tau {\sigma'}_t} \bigg)  \bigg] + c \,, 
\label{eq:RECO}
\end{equation}

\begin{figure}
\begin{centering}
\includegraphics[width=0.99\columnwidth]{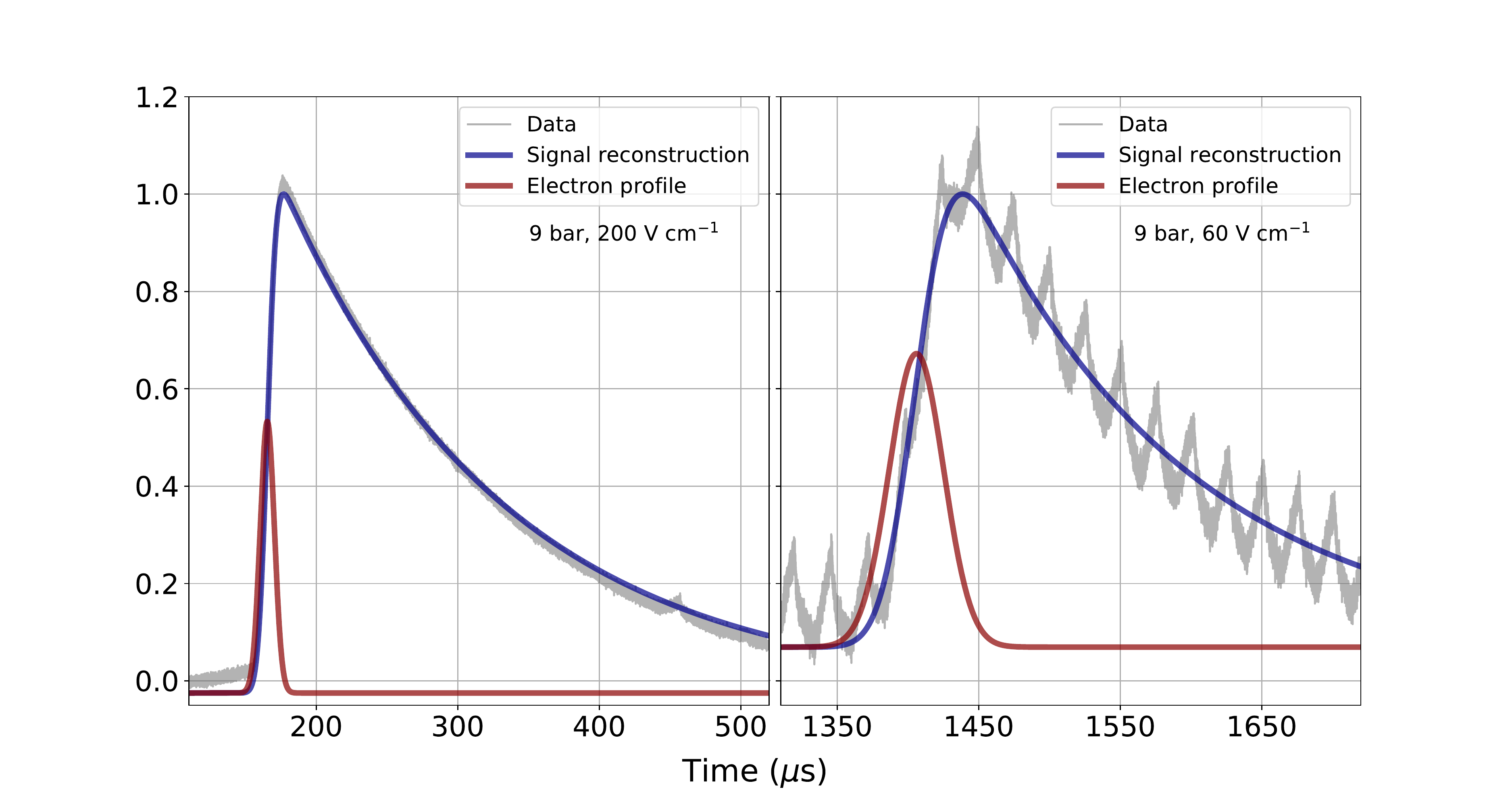}
\par\end{centering}
\caption{Fit to mean waveforms at different electric field strengths. The gray line shows the mean waveform, the blue line shows the reconstructed signal from fitting Eq. \ref{eq:RECO}, and the red line shows the electron swarm profile extracted from the reconstruction. The waveforms peak are normalized to 1 so the fit can be seen for the larger (left) and smaller (right) waveforms observed.
\label{fig:waveformfit}}
\end{figure}

We fit this function to averages of 1000 waveforms to extract drift parameters of interest $t_0$, $\sigma_t$, $\tau$ and nuisance parameters $a$ and $c$. Example fits for a particularly narrow and particularly wide pulse, as well as the implied arriving electron distribution from those fits, are shown in Fig.~\ref{fig:waveformfit}. For the smallest amplitude waveforms at low reduced field, the residual power supply switching noise becomes visible over background, this is shown in  Fig.~\ref{fig:waveformfit} right.  To establish that this did not bias the pulse shape extraction, fits with different flash lamp brightness at low fields  were compared, effectively varying the relative signal to noise ratio. No significant changes in fit parameters was observed.

The decay constant $\tau$ was measured in vacuum runs and fixed to this reference value in all subsequent fits. The other parameter values were seeded using approximate measures of the average pulse and then allowed to float freely in the fit.  Both the anode and cathode waveforms are fit, and the total pulse-to-pulse time $t_{pp}$ is extracted by subtraction.  To obtain $t_d$ from this measurement, it is necessary to apply end-effect corrections including accounting for the inductive width effect, described below, that typically give one percent level effects.

Electrons cross the collection gap very quickly, because it is short (1.6~mm) and has high E-field ($\geq$ 3000~V/cm). The extraction gap, on the other hand, is larger and has a less intense electric field, and so the inductive width $w$ must be accounted for.  In the extraction gap, which is of length $d_{gap}=3.9$~mm there is a fixed electric field of 300 V/cm for all runs.  To account for this region, at each pressure we first analyze the 300 V/cm drift-field data, where both extraction and drift regions have the same field strength. The total drift distance is then $d+d_{gap}$, with uniform electric field, and electron travel time $t_{pp}$. This allows for a reliable measurement of both $v_{d,300}$ and $D_{L,300}$.  For each lower drift-field, half the transit time across the extraction region $w=t_{gap}=d_{gap}/v_{d,300}$ is subtracted from $t_{pp}$ to yield the drift time $t_d$, with the drift distance restricted only to $d$. This provides a measurement of drift velocity only in the lower field region. Similarly, the diffusion taking place in the extraction region is subtracted in quadrature from the final fitted $\sigma_t$. The latter correction is generally found to be negligible, since $\sigma_t$ scales as $\sqrt{t}$, and the field in the extraction region is larger than in the drift region which implies electrons spend little time there.

After this procedure, the drift velocity $v_d$ and longitudinal diffusion constant $D_L$ have been extracted for every pressure and electric field configuration. Theoretically, a scaling is expected with $v_d$ as a universal function of $E/P$ and  $P D_L$ as a universal function of $E/P$, for all $E$ and all $P$.  Our data at each $E$ for each $P$ allow to test the scaling with $E/P$ and also extract these universal functions.

\subsection{Systematic Uncertainties}
\label{sec:Systematics}

The sources of systematic uncertainty on the velocities and diffusion coefficients derive from experimental uncertainty on the drift time and width of the electron swarm profile (the latter of which factors in the calculation of the diffusion coefficient at cubic power), as well as the uncertainty on the gas mixture fraction, drift distance, and absolute gas pressure.

The measurement of the variance, $\sigma_t^2$, and the drift time, $t_d$, of the electron profile are affected by the precision of the oscilloscope used to record the waveforms, the reconstruction of the electron swarm via fitting the waveform, and possible non-uniformities in the electric field, and any residual effects from space charge broadening. The uncertainty on the estimation of the variance and drift time from the waveform reconstruction is related to the accuracy of the fitting function, discussed above, and is found to be sub-dominant to other sources of uncertainty. Space charge effects are small at these flash lamp intensities, but not absolutely negligible.  We constrained the effects of spacecharge on the measurement by varying the lamp brightness and repeating the fit at the highest pressure (10 bar) and lowest electric field (50~V/cm), where space charge effects will be maximally severe. A small correlated variation was observed at the level of  1.5\% in the pulse width and 0.12\% in the drift time.  To study the effects of fringe fields and electron transit time at the induction region, the collection field was varied over the range 2000 to 5000~V/cm, and a variation of the pulse width was observed at the 2.7\% and in the drift time at the  0.2\% level.  No uncertainty on the drift parameters appeared to be caused by the extraction field, after applying the end-effect correction.  The overall uncertainty on the parameters is determined by the above effects added in quadrature and is found to be $\delta \sigma_t^2/\sigma_t^2 = 0.062$ on the squared pulse width and $\delta t_d / t_d = 0.0046$ on the drift time.

The measurement of the fraction of each gas in the mixtures is primarily affected by the precision of the transducer pressure gauge. The uncertainty on the transducer pressure measurement is estimated to be 0.01 bar. This gives an uncertainty on the fraction of helium in the mixtures on the order of 1\%.
The uncertainty on the measurement of the drift distance is estimated to be less than 1\% (d~=~14.1~$\pm$~0.1~cm), and the uncertainty on the absolute pressure of the system is estimated to be 0.1 bar. This gives a 10\% pressure uncertainty at 1 bar and a 1\% pressure uncertainty at 9 bar.

\begin{figure}[b!]
\begin{centering}
\includegraphics[width=0.5\columnwidth]{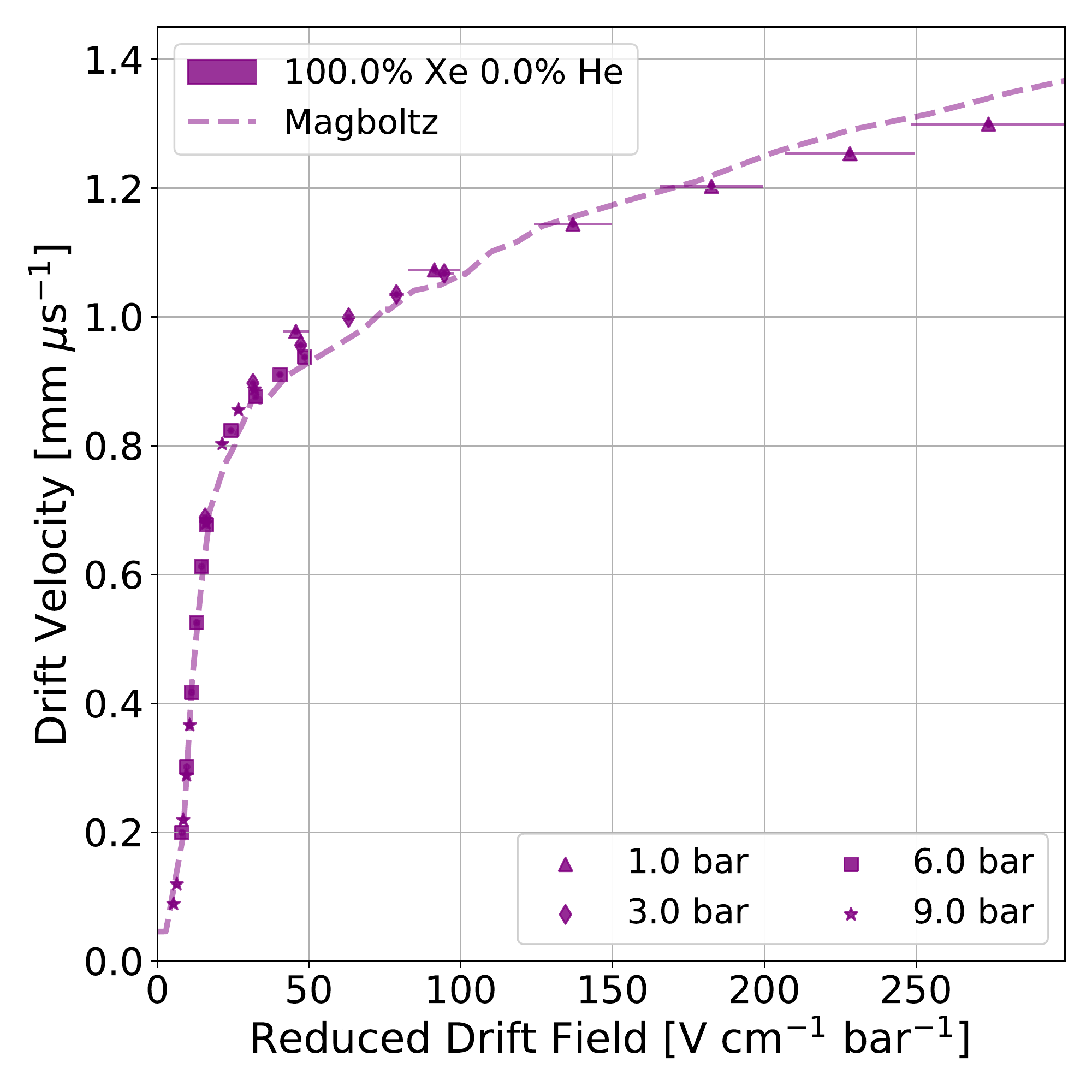}
\includegraphics[width=0.99\columnwidth]{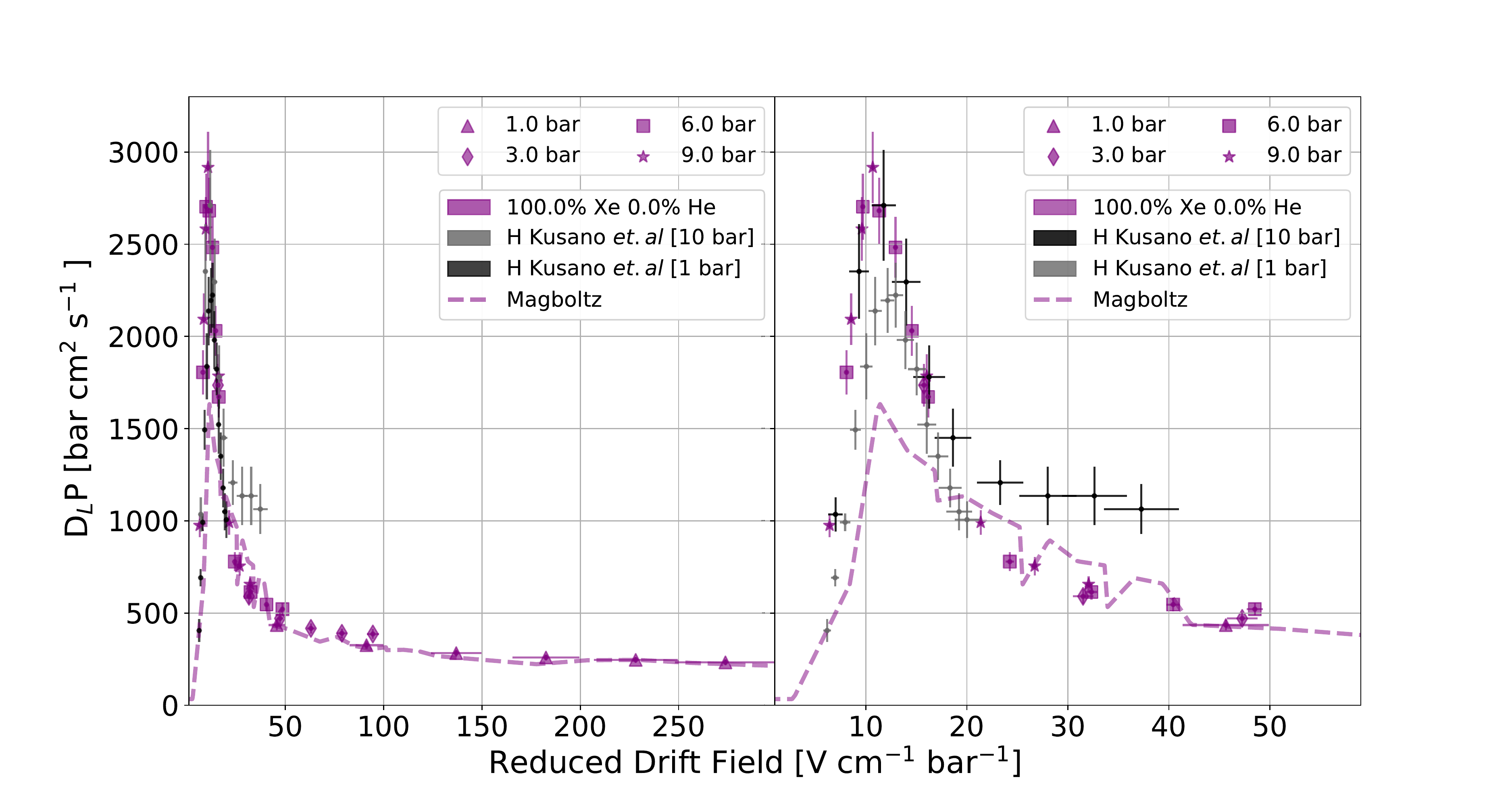}

\par\end{centering}
\caption{Top: Drift velocity in pure xenon. Bottom: Longitudinal diffusion in xenon. Right panel shows a closeup of the region of most interest to TPC experiments such as NEXT.\label{fig:xepure}}
\end{figure}

\section{Results}\label{sec:Results}
\subsection{Pure Xenon Drift Velocity and Diffusion}\label{sec:pure}

We first studied pure xenon gas. Data were taken at pressures from 1 to 9~bar and electric fields from 50 to 300~V/cm. The data were collected and analyzed according to the protocols outlined in the previous sections.   The measured drift velocity is in very good agreement with theoretical predictions from {\tt MagBoltz}~\cite{MagBoltz}(Version 11.8) \footnote{This version of {\tt MagBoltz} accounts for the correlation length of gas mixtures, and at the time of writing was not public.} over the full range, as shown in Fig.~\ref{fig:xepure}, top. Notably we observe the expected scaling behaviour across all pressures, with $v_d$ exhibiting universality as a function of the scaling parameter $E/P$ for all $E$ and all $P$.  The measured diffusion constant $D_L$ also shows the expected scaling as $P\cdot D_L$. Our data are consistent at low $E/P$ with previous measurements in high pressure xenon gas from Kusano {\em et al.}~\cite{Kusano:2013uwa}.  However, the measurements diverge from {\tt MagBoltz} predictions at mid to low $E/P$, demonstrating a much stronger rise.  Our data, however, match much more closely with {\tt MagBoltz} at higher $E/P$  ($\geq$ 20 V/cm-bar) than those from \cite{Kusano:2013uwa}. Data from Pack {\em et al.}~\cite{pack1992longitudinal} was also considered however due to the large systematic uncertainties a comparison could not be made.

\begin{figure}
\begin{centering}
\includegraphics[width=0.99\columnwidth]{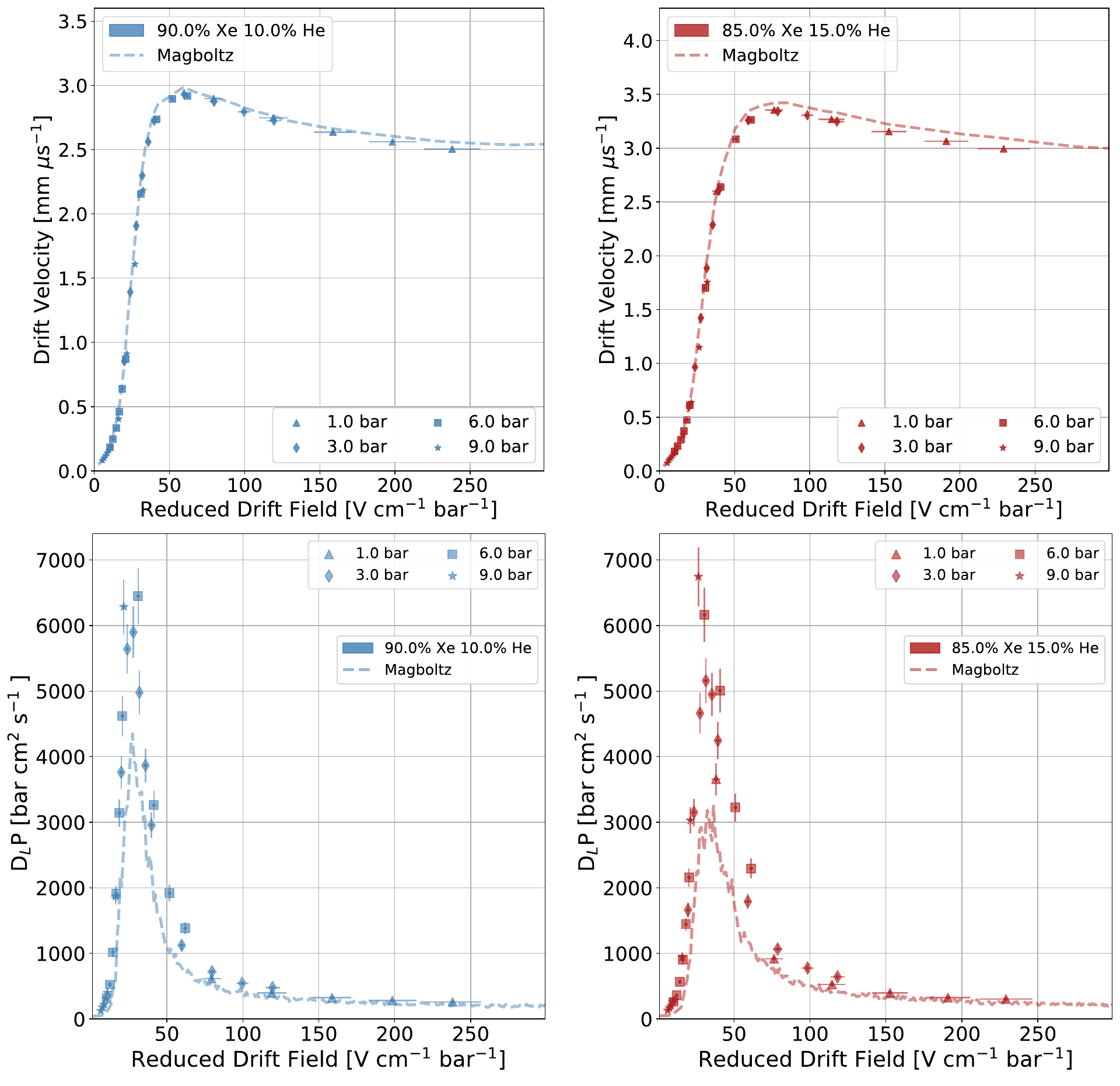}
\par\end{centering}
\caption{Top: Drift velocity in xenon-helium mixtures. Bottom: Longitudinal diffusion in xenon-helium mixtures.\label{fig:xeheplots}}
\end{figure}

\subsection{Xenon-Helium Mixtures}\label{sec:xehe}

Data were taken in gas mixtures of 10\% and 15\% helium in xenon at pressures from 1 to 9 bar and electric fields from 50 to 300 V/cm following the gas mixing and data taking protocols defined in previous sections.  The measured velocities and diffusion coefficients are shown in Fig.~\ref{fig:xeheplots}.  Once again, strong scaling of $v_d$ and $P\cdot D_L$ with $E/P$ is observed. Excellent agreement with drift velocity predictions from {\tt MagBoltz} is also observed.  Again, a notable excess in longitudinal diffusion relative to {\tt MagBoltz} is present, this time at higher values of $E/P$ than pure xenon. These values of $E/P$ are within the range relevant to TPC experiments that may use this gas mixture, and we give more detailed comments on the implications below.

\subsection{Discussion}\label{sec:Discussion}

We have presented data taken using an electron drift chamber which allows us to measure the drift velocity and longitudinal diffusion coefficients of electrons in xenon gas and in xenon-helium mixtures that are of interest as diffusion reducing agents at concentrations of 10\% and 15\% helium by volume.  In all cases we find excellent agreement with predictions of velocity from the {\tt MagBoltz} software. However, we find that longitudinal diffusion is under-predicted at very low $E/P$ in xenon, and at higher $E/P$ in mixtures with helium, for all relevant $E$ and $P$.  
That such a phenomenon at a higher $E/P$ in xenon-helium mixtures than in pure xenon may be expected, since helium has the effect of cooling the drift electrons, leading them to have lower average energies resulting in the same impact on the electron-xenon cross section at lower $E/P$. Which would result in the xenon excess shifting to the higher $E/P$ in xenon-helium mixtures. Thus any imperfection in the low-energy xenon-electron cross section model within {\tt MagBoltz} would be expected to exhibit such a behaviour.  These data may suggest a need to revisit the electron-xenon cross section model in the present generation of simulation tools.
\begin{figure}
\begin{centering}
\includegraphics[width=0.99\columnwidth]{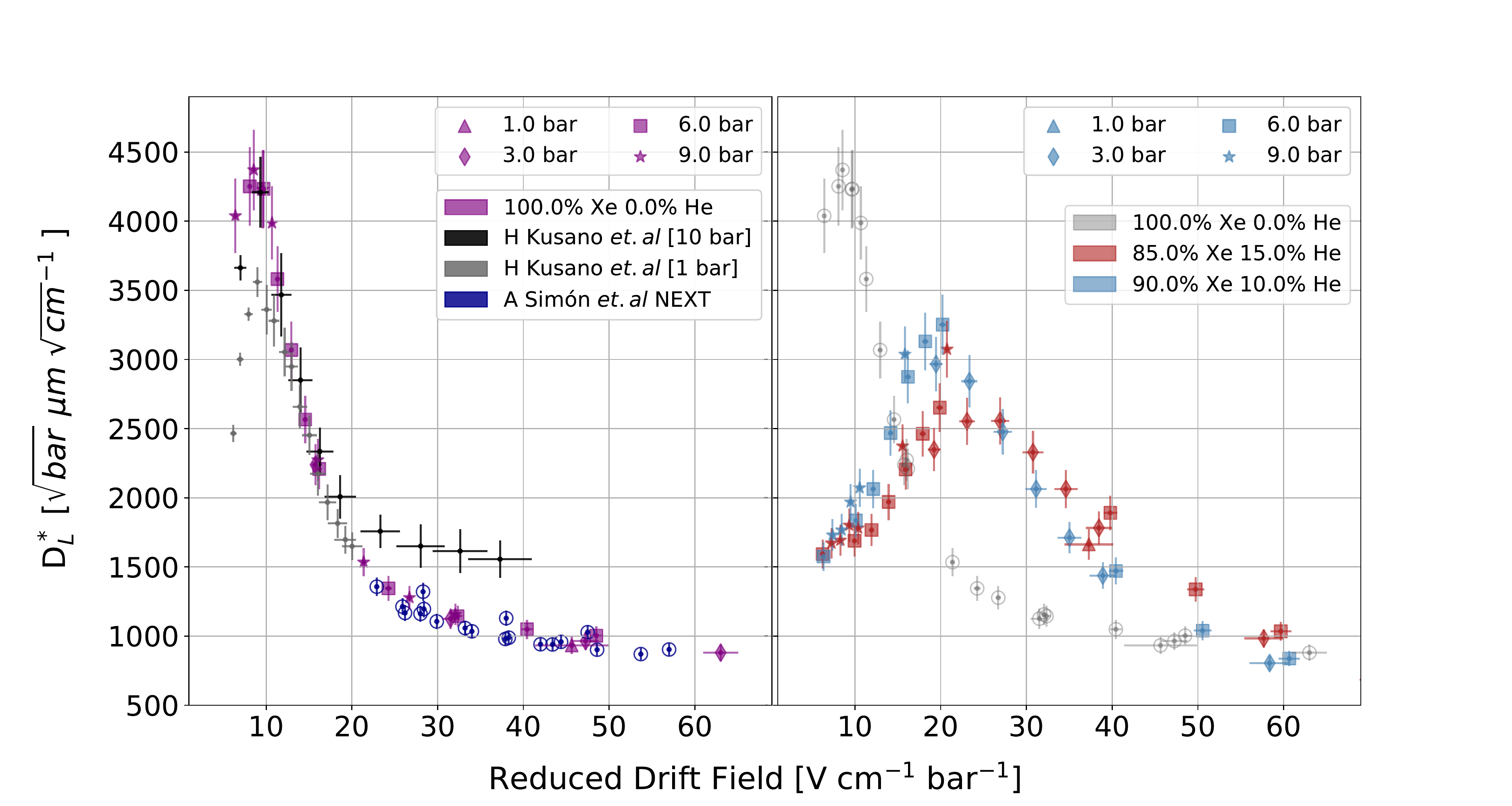}
\par\end{centering}
\caption{Comparison of $D_L^*$ between pure xenon (left) and xenon-helium mixtures (right).\label{fig:TPCDiffusion}}
\end{figure}

Typical TPC operating fields are in the range 300-500~V/cm and the operating pressures for the NEXT program are in the range 10-15 bar.  This implies a relevant range of $E/P$ of 20 to 50 V/cm-bar for TPC detectors.  It appears the longitudinal diffusion in this regime will be larger than has been predicted in the past, and larger with a helium in xenon mixture than in pure xenon. On the other hand, consideration of Fig.~\ref{fig:xeheplots} alone does not provide a full account of the implications for experiments.  While part of the effect on spatial resolution from an electron-cooling additive can derive from a change in $D_L$, further effects must be given consideration in terms of detector performance. For example, not only is $D_L$ changed by the addition of helium, but so is $v_d$. The much faster electrons will naturally diffuse less, since the diffusion coefficient specifies the dependence of the diffusion process upon time, whereas the electrons have to travel a finite distance. The drift velocity is enhanced in xenon-helium mixtures, which to some extent counteracts an increase in $D_L$.  A useful quantity in this regard is the ``TPC diffusion'' $D_L^*$ which encompasses the expected spread of charge that has travelled a fixed distance: where $v_d$ is the drift velocity, $T^0$ is the temperature that is extrapolated to (293.15$K$), $D_L$ is the diffusion coefficient with $T$ and $P$ being the temperature and pressure it was measured at.

\begin{equation} \label{Dstar}
D_L^*=\sqrt{\frac{T^0}{T}\frac{2 P D_L}{v_d}}
\end{equation}

Comparing values of $D_L^*$ is a clearer way to establish the relative spatial resolution of different gas mixtures. We show a comparison of our data on pure xenon to xenon-helium mixtures in Fig.~\ref{fig:TPCDiffusion}. Also shown in this plot are previous data from the NEXT experiment operated with pure xenon, from \cite{Simon:2018vep}, which was reported in similar units.  Our data match very well with these observations at higher $E/P$ and, as already noted, with data from Kusano {\em et al}. at lower $E/P$. A comparison of the reduced longitudinal diffusion $D_L^*$ in the TPC operating regime of interest suggests that the addition of helium at the 10-15\% level will increase the scale of longitudinal diffusion by around 50\%. 

\begin{figure}
\begin{centering}
\includegraphics[width=0.99\columnwidth]{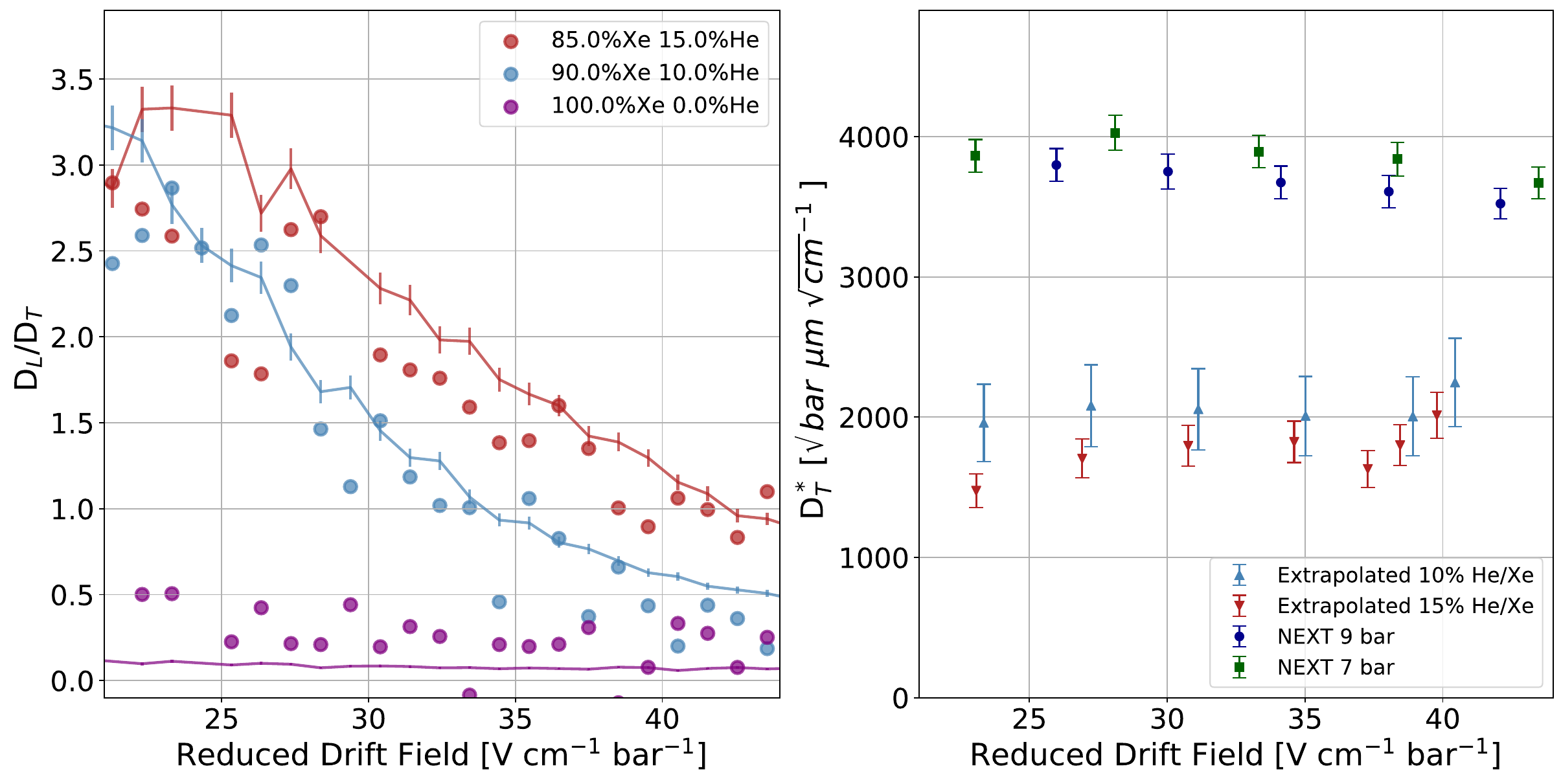}
\par\end{centering}
\caption{Left: validation of the Wannier Relation at low fields using {\tt MagBoltz} simulation. Dots show $\frac{1}{v_d}\frac{E}{P}\frac{d v_d}{d(E/P)}$ and lines show $D_L / D_T$ calculated from {\tt MagBoltz} output, which are equivalent if this relation holds. The gases shown are 0\%, 10\% and 15\% helium in xenon; Right: extrapolated transverse diffusion utilizing the measurements of $D_L$ and $v_d$ from this work, assuming validity of the Wannier relation, compared to data from NEXT in pure xenon. Helium additives may offer a factor of $\sim$2 improvement in terms of transverse diffusion control. \label{fig:DTExtrap}}
\end{figure}

Notably we have not, in this work, measured a transverse diffusion $D_T$, because our apparatus is not well equipped for this task.  However, the transverse and longitudinal diffusion coefficients are theoretically related to each other by the Wannier relation of Eq. \ref{eq:WannierRel}. Given that the right hand side of this relation is a function of $v_d$ and $E/P$ only, and these match very well between our data and simulation, we might expect discrepancies in $D_L$ to be mirrored by similar ones in $D_T$ in order to maintain a constant ratio, wherever this relation is applicable.  While it is clear that the Wannier relation is not universally applicable in these gases - in particular at high fields the negative $dv_d/d(E/P)$ would suggest a negative and hence unphysical value for $D_T$. {\tt MagBoltz} simulations verify that the Wannier relation is expected to apply in the immediate region of interest, 20-50 V/cm-bar in Xe/He mixtures. In pure Xe, however, $D_L$ is so small relative to $D_T$ that the Wannier relation cannot be verified with {\tt MagBoltz}, since the accuracy required on the slope of the mobility is less the uncertainty.  We quantify this uncertainty by evaluating the ($\frac{D_L}{D_T}$) and ($\frac{1}{v_d}\frac{E}{P}$) independently with {\tt MagBoltz} at each reduced drift field (Fig.~\ref{fig:DTExtrap}, left.), and calculating the root-mean-square between them. This provides the accuracy with which  {\tt MagBoltz} verifies the applicability of the relation, which is taken as the uncertainty on the extrapolation of $D_T$ from measured values of $D_L$.  We find that this extrapolation is validated to better than 5\% in Xe/He mixtures, but not validatable to better than  200\% in pure Xe. Since meaningful extrapolation of $D_T$ from $D_L$ could not be verified for pure xenon, in this work we present only the extrapolated $D_T$ values for the mixtures.  The results are shown in Fig.~\ref{fig:DTExtrap}, right.  The error bars include the measured errors on $D_L$ as well as the uncertainty on the Wannier relation in this regime. As apparent in Fig.~\ref{fig:MagboltzPredictions}, the predicted improvement in diffusion reduction is much larger in $D_T$ than in $D_L$, and so even with a 50\% increase over predictions of diffusion in the $E/P$ range of interest, a substantial improvement in transverse diffusion is still apparent.  Experimental verification of this extrapolation will be an important subject for future work.

\section{Conclusion}

We have measured the drift velocity and longitudinal diffusion coefficient in pure xenon and in xenon-helium mixtures at pressures between 1 and 9 bar and electric fields between 50 and 300~V/cm. We observe the expected scaling with $E/P$ in all distributions, and measured drift velocities are in excellent agreement with predictions from {\tt MagBoltz}.  Measured diffusion coefficients in pure xenon are in excellent agreement with world data at all $E/P$ and with {\tt MagBoltz} at high $E/P$.  At low $E/P$ we observed larger diffusion than had been predicted, suggesting the low energy electron-xenon cross section may not be properly estimated.  We observe excellent agreement in drift velocity for the xenon-helium system at 10\% and 15\% concentrations as well. In this case a longitudinal diffusion discrepancy occurs at higher $E/P$, understood to happen through the cooling effect of the helium on the drifting electrons.  Although longitudinal diffusion $D_L^*$ is larger than expected by around 50\% and larger than in pure xenon, the transverse diffusion remains to be studied experimentally.  Given the scale of predicted $D_T$ reduction by electron cooling and the presented measurements of $D_L$, we may expect that spatial resolution improvements of a factor of $\sim$2 may be derived from xenon-helium gas mixtures in future experiments.

\section*{Acknowledgement}
We would like to think Steve Biagi for his insightful discussions and his assistance with {\tt MagBoltz}. We would also like to think Dami\'an Garc\'ia and Lorenzo Mu\~niz for their fruitful discussions.

The electron drift system described in this paper was developed as a purity monitor for the Barium Tagging R\&D System at UTA. The work described was supported by the Department of Energy under Early Career Award number {DE-SC0019054}. The University of Texas at Arlington group is also supported by Department of Energy Award {DE-SC0019223}.  DGD is supported by MINECO (Spain) under the Ramon y Cajal program (contract RYC-2015-18820).

The NEXT Collaboration acknowledges support from the following agencies and institutions: the European Research Council (ERC) under the Advanced Grant 339787-NEXT; the European Union’s Framework Programme for Research and Innovation Horizon 2020 (2014-2020) under the Marie Skłodowska-Curie Grant Agreements No. 674896, 690575 and 740055; the Ministerio de Econom\'ia y Competitividad of Spain under grants FIS2014-53371-C04, the Severo Ochoa Program SEV-2014-0398 and the Mar\'ia de Maetzu Program MDM-2016-0692; the GVA of Spain under grants PROMETEO/2016/120 and SEJI/2017/011; the Portuguese FCT under project PTDC/FIS-NUC/2525/2014, under project UID/FIS/04559/2013 to fund the activities of LIBPhys, and under grants PD/BD/105921/2014, SFRH/BPD/109180/2015 and SFRH/BPD/76842/2011. Finally, we are grateful to the Laboratorio Subterr\'aneo de Canfranc for hosting and supporting the NEXT experiment.

\bibliography{main}

\begin{thebibliography}{10}

\bibitem{Gonzalez-Diaz:2017gxo}
D.~Gonzalez-Diaz, F.~Monrabal, and S.~Murphy.
\newblock {Gaseous and dual-phase time projection chambers for imaging rare
  processes}.
\newblock {\em Nucl. Instrum. Meth.}, A878:200--255, 2018.

\bibitem{alvarez2013initial}
{V {\'A}lvarez, et al. (The NEXT collaboratio)}.
\newblock Initial results of next-demo, a large-scale prototype of the next-100
  experiment.
\newblock {\em JINST}, 8(04):P04002, 2013.

\bibitem{alvarez2013operation}
{V {\'A}lvarez, et al. (The NEXT collaboratio)}.
\newblock Operation and first results of the next-demo prototype using a
  silicon photomultiplier tracking array.
\newblock {\em JINST}, 8(09):P09011, 2013.

\bibitem{alvarez2013near}
{V {\'A}lvarez, et al. (The NEXT collaboratio)}.
\newblock Near-intrinsic energy resolution for 30--662 kev gamma rays in a high
  pressure xenon electroluminescent tpc.
\newblock {\em Nucl. Instrum. Meth.}, 708:101--114, 2013.

\bibitem{Ferrario:2015kta}
{P Ferrario, et al. (The NEXT collaboratio)}.
\newblock {First proof of topological signature in the high pressure xenon gas
  TPC with electroluminescence amplification for the NEXT experiment}.
\newblock {\em JHEP}, 01:104, 2016.

\bibitem{Renner:2018ttw}
{J Renner, et al. (The NEXT collaboratio)}.
\newblock {Initial results on energy resolution of the NEXT-White detector}.
\newblock {\em JINST}, 13(10):P10020, 2018.

\bibitem{henriques2019electroluminescence}
{CAO Henriques, et al. (The NEXT collaboratio)}.
\newblock Electroluminescence tpcs at the thermal diffusion limit.
\newblock {\em JHEP}, 2019(1):27, 2019.

\bibitem{henriques2017secondary}
{CAO Henriques, et al. (The NEXT collaboratio)}.
\newblock Secondary scintillation yield of xenon with sub-percent levels of co2
  additive for rare-event detection.
\newblock {\em Physics Letters B}, 773:663--671, 2017.

\bibitem{pack1962drift}
JL~Pack, RE~Voshall, and AV~Phelps.
\newblock Drift velocities of slow electrons in krypton, xenon, deuterium,
  carbon monoxide, carbon dioxide, water vapor, nitrous oxide, and ammonia.
\newblock {\em Physical Review}, 127(6):2084, 1962.

\bibitem{pack1992longitudinal}
JL~Pack, RE~Voshall, AV~Phelps, and LE~Kline.
\newblock Longitudinal electron diffusion coefficients in gases: Noble gases.
\newblock {\em Journal of applied physics}, 71(11):5363--5371, 1992.

\bibitem{koizumi1986momentum}
T~Koizumi, E~Shirakawa, and I~Ogawa.
\newblock Momentum transfer cross sections for low-energy electrons in krypton
  and xenon from characteristic energies.
\newblock {\em Journal of Physics B: Atomic and Molecular Physics},
  19(15):2331, 1986.

\bibitem{bowe1960drift}
JC~Bowe.
\newblock Drift velocity of electrons in nitrogen, helium, neon, argon,
  krypton, and xenon.
\newblock {\em Physical Review}, 117(6):1411, 1960.

\bibitem{hunter1988low}
SR~Hunter, JG~Carter, and LG~Christophorou.
\newblock Low-energy electron drift and scattering in krypton and xenon.
\newblock {\em Physical Review A}, 38(11):5539, 1988.

\bibitem{alvarez2013ionization}
{V {\'A}lvarez, et al. (The NEXT collaboratio)}.
\newblock Ionization and scintillation response of high-pressure xenon gas to
  alpha particles.
\newblock {\em JINST}, 8(05):P05025, 2013.

\bibitem{lorca2014characterisation}
{D Lorcia, et al. (The NEXT collaboratio)}.
\newblock Characterisation of next-demo using xenon k$\alpha$ x-rays.
\newblock {\em JINST}, 9(10):P10007, 2014.

\bibitem{Simon:2018vep}
{A Sim{\'o}n, et al. (The NEXT collaboratio)}.
\newblock {Electron drift properties in high pressure gaseous xenon}.
\newblock {\em JINST}, 13(07):P07013, 2018.

\bibitem{patrick1991electron}
EL~Patrick, ML~Andrews, and A~Garscadden.
\newblock Electron drift velocities in xenon and xenon-nitrogen gas mixtures.
\newblock {\em Applied physics letters}, 59(25):3239--3240, 1991.

\bibitem{english1953grid}
WN~English and GC~Hanna.
\newblock Grid ionization chamber measurements of electron drift velocities in
  gas mixtures.
\newblock {\em Canadian Journal of Physics}, 31(5):768--797, 1953.

\bibitem{kobayashi2006ratio}
Shingo Kobayashi, Nobuyuki Hasebe, Takehiro Hosojima, et~al.
\newblock Ratio of transverse diffusion coefficient to mobility of electrons in
  high-pressure xenon and xenon doped with hydrogen.
\newblock {\em Japanese journal of applied physics}, 45(10R):7894, 2006.

\bibitem{Gonzalez-Diaz:2015oba}
{D Gonz{\'a}lez-D{\'\i}az, et al. (The NEXT collaboratio)}.
\newblock {Accurate $\gamma$ and MeV-electron track reconstruction with an
  ultra-low diffusion Xenon/TMA TPC at 10 atm}.
\newblock {\em Nucl. Instrum. Meth.}, A804:8--24, 2015.

\bibitem{Nakajima:2015meb}
{Y Nakajima, et al. (The NEXT collaboratio)}.
\newblock {Measurement of scintillation and ionization yield with high-pressure
  gaseous mixtures of Xe and TMA for improved neutrinoless double beta decay
  and dark matter searches}.
\newblock {\em JINST}, 11(03):C03041, 2016.

\bibitem{XePa}
B.~J.~P Jones.
\newblock {XePA Project: Drift properties of helium added to xenon at 10 bar}.
\newblock \url{http://www-hep.uta.edu/~bjones/XePA/}, 2016.

\bibitem{Felkai:2017oeq}
{R Felkai, et al. (The NEXT collaboratio)}.
\newblock {Helium-Xenon mixtures to improve the topological signature in high
  pressure gas xenon TPCs}.
\newblock {\em Nucl. Instrum. Meth.}, A905:82--90, 2018.

\bibitem{Li:2015rqa}
Yichen Li, Thomas Tsang, Craig Thorn, et~al.
\newblock {Measurement of Longitudinal Electron Diffusion in Liquid Argon}.
\newblock {\em Nucl. Instrum. Meth.}, A816:160--170, 2016.

\bibitem{coelho2007measurement}
LCC Coelho, HMNBL Ferreira, JAM Lopes, et~al.
\newblock Measurement of the photoelectron-collection efficiency in noble gases
  and methane.
\newblock {\em Nucl. Instrum. Meth.}, 581(1-2):190--193, 2007.

\bibitem{borghesani2011injection}
AF~Borghesani and P~Lamp.
\newblock Injection of photoelectrons into dense argon gas.
\newblock {\em Plasma Sources Science and Technology}, 20(3):034001, 2011.

\bibitem{smejtek1973hot}
Pavel Smejtek, M~Silver, KS~Dy, and David~G Onn.
\newblock Hot electron injection into dense argon, nitrogen, and hydrogen.
\newblock {\em The Journal of Chemical Physics}, 59(3):1374--1384, 1973.

\bibitem{ViehlandBook}
Larry~A. Viehland.
\newblock {\em Gaseous Ion Mobility, Diffusion, and Reaction}.
\newblock Number~36. Springer Nature Switzerland, 2018.

\bibitem{robson1972thermodynamic}
Robert~Edward Robson.
\newblock A thermodynamic treatment of anisotropic diffusion in an electric
  field.
\newblock {\em Australian Journal of Physics}, 25(6):685--694, 1972.

\bibitem{MagBoltz}
{Stephen Biagi}.
\newblock Magboltz - transport of electrons in gas mixtures.
\newblock \url{http://magboltz.web.cern.ch/magboltz/}, 2018.
\newblock [Online; accessed October-2018].

\bibitem{Kusano:2013uwa}
H.~Kusano, J.~A.~M. Lopes, M.~Miyajima, and N.~Hasebe.
\newblock {Longitudinal and transverse diffusion of electrons in high-pressure
  xenon}.
\newblock {\em JINST}, 8:C01028, 2013.

\end{thebibliography}
\bibliographystyle{unsrt}
\end{document}